\documentstyle[aps,bezier,epsfig,amsmath,amssymb]{revtex}
\begin{document}

\newcommand{\be}{\begin{eqnarray}}
\newcommand{\ee}{\end{eqnarray}}
\newcommand{\beq}{\begin{equation}}
\newcommand{\eeq}{\end{equation}}
\newcommand{\xx}{\begin{eqnarray*}}
\newcommand{\yy}{\end{eqnarray*}}
\newcommand{\nn}{\nonumber}
\newcommand{\Vol}{{\rm Vol}}
\newcommand{\sign}{{\rm sign}}
\newcommand{\tr}{{\rm Tr}}
\def\cut{{}\hfill\cr \hfill{}}

\title{Reply to Comment on `` Universal Fluctuations in Correlated Systems'', 
 by B. Zheng and S. Trimper, cond-mat/0109003. }

\author{ S.T. Bramwell$^1$,K. Christensen{$^2$}, J.-Y. Fortin$^{3}$, P.C.W.
Holdsworth$^4$,
H.J. Jensen$^5$, S. Lise{$^5$}, J.M. L\'opez{$^{6}$}, M. Nicodemi{$^{5,7}$}, J.-F.
Pinton$^4$, M. Sellitto$^4$}

\address{$^1$ Department of Chemistry, University College London, 20
Gordon
Street, London, WC1H~0AJ, United Kingdom. \\
{$^2$}Blackett Laboratory, Imperial College,
Prince Consort Road, London SW7 2BZ United Kingdom.\\
$^3$ CNRS, UMR 7085, Laboratoire de Physique Theorique, Universite 
Louis Pasteur, 67084 Strasbourg, France. \\
$^4$ Laboratoire de Physique, Ecole Normale Sup\'erieure,
46 All\'ee d'Italie, F-69364 Lyon cedex 07, France.\\
$^5$ Department of Mathematics, Imperial College London, London, SW7,United Kingdom\\.
$^6$ Instituto de Fisica de Cantabria, CSIC-UC,
Santander 39005, Spain.\\
$7$ Dipartimento di Fisica, Universit\`a ``Federico II'', INFM, INFN, Napoli, Italy}

\maketitle

\vspace{1cm}

\noindent{\bf To appear in {\it Phys. Rev. Lett.}}

\vspace{1 cm}

\medskip

\medskip

\noindent{\Large\bf Bramwell et al. reply}

\medskip

\medskip

Zheng and Trimper confirm our conjecture in paper~\cite{1}, that
the probability distribution for order parameter fluctuations in the 2D and 3D
Ising models at a temperature $T^{\ast}(L)$ slightly below $T_C (L
\rightarrow \infty)$ approximates
the universal functional form  of the 2D-XY model in its low temperature
phase \cite{2}.
They show quantitatively that $T_C-T^{\ast}(L)$ scales as $L^{-1/\nu}$.
The XY-type scaling is, of course, only one locus
in the ${L^{-1},T}$ plane and for general ${L^{-1},T}$  the PDF is not of the
XY form \cite{1,3}. The point of departure between our interpretation of
this result and that of
Zheng and Trimper is that we attribute the PDF to critical fluctuations and
they do not.
We are pleased to take the opportunity to discuss this point in detail.

The 2D-XY model is critical throughout the low temperature regime
with diverging longitudinal fluctuations.
It is therefore incorrect to think that critical
``fluctuation are mainly rotational''.
The physics of a phase transition in a spin system with continuous symmetry 
is controlled by the divergence of the longitudinal susceptibility;
the transverse susceptibility being infinite at all temperatures. 
The lengthening magnetization
vector as order develops drives the diffusion constant around the 
circle (in the XY model) to zero in the thermodynamic limit, and
consequently rotational symmetry is broken.
The scalar magnetization is therefore a critical quantity, as can be seen
through any finite-size scaling criterion. 
However, it is rather a special
limit for critical fluctuations: despite the susceptibility diverging
as $\sim L^{d-\eta}$ and $\sigma/\langle m\rangle$ remaining independent of
system size, the latter ratio is small,  $\approx \eta/4$ \cite{2}.
The result, paradoxically, is that the
divergent fluctuations never bring the order parameter near
the limits $m = 0$ and $m = 1$. The critical fluctuations therefore occur
without ever changing the fixed topology (or symmetry)
imposed by the corrections to the thermodynamic limit~\cite{2}.
That is, there is a barrier to jump to arrive at $m=0$, but 
$\langle m \rangle \sim L^{-\eta/2}$ is not an intensive variable and the free
energy barrier is not extensive; it is a correction to the
thermodynamic limit and a pure effect of criticality.

Zheng and Trimper find that, for the Ising model,
the measured correlation length at $T^{\ast}(L)$ is small compared with
$L$. This property should ensure that fluctuations in the
Ising systems studied can be described in a similar way to those of the XY
model.
Indeed the authors point out that at $T^{\ast}(L)$,
the order parameter remains far from the minimum of probability: $m = 0$.
However they are wrong to conclude that the fluctuations at this
temperature are ``not a characteristic property at the critical point'':
the observation of the universal fluctuations at constant $s = L \tau$
means that the ``small'' correlation length is fixed by the system size;
it does diverge in the
thermodynamic limit and the observed property is a critical property. If
the correlation length remained finite, (for example if
the transition were first order) the central limit theorem would
apply and the limit distribution would be Gaussian, or a closely related
function \cite{4}.

The critical point is a singular point in the thermodynamic limit and
admits many definitions in a corresponding finite system. The existence of
the family of loci $[T(L),L^{-1}]$, all collapsing onto $T_C$ in the
thermodynamic limit has been addressed in  ~\cite{3,4,5}. However, in  view
of the fact that many complex systems approximate the XY functional form,
it seems that the important question for the Ising model is: does the locus
of points identified here have any special properties, or is it just one of
many curves?

\end{document}